\documentclass{ws-ijgmmp}

\begin{document}

\def\be{\begin{equation}}
\def\ee{\end{equation}}
\def\bea{\begin{eqnarray}}
\def\eea{\end{eqnarray}}
\def\tr{\mathrm{tr}\, }
\def\nn{\nonumber \\}
\def\e{\mathrm{e}}

\markboth{Shin'ichi Nojiri and Sergei D. Odintsov}
{Accelerating cosmology in modified gravity}

%%%%%%%%%%%%%%%%%%%%% Publisher's Area please ignore %%%%%%%%%%%%%%%
%
\catchline{}{}{}{}{}
%
%%%%%%%%%%%%%%%%%%%%%%%%%%%%%%%%%%%%%%%%%%%%%%%%%%%%%%%%%%%%%%%%%%%%

\title{Accelerating cosmology in
modified gravity: from convenient $F(R)$ or string-inspired theory to
bimetric $F(R)$ gravity %\footnote{************************.}
}

\author{SHIN'ICHI NOJIRI%\footnote{******************.}
}

\address{Department of Physics, Nagoya University, Nagoya
464-8602, Japan \\
Kobayashi-Maskawa Institute for the Origin of Particles and
the Universe, Nagoya University, Nagoya 464-8602, Japan \\
%\,\footnote{*********************.}\\
\email{nojiri@phys.nagoya-u.ac.jp%\footnote{***************}
} }

\author{SERGEI D. ODINTSOV}

\address{Consejo Superior de Investigaciones Cient\'{\i}ficas,
Institut de Ciencies de l'Espai (ICE),
(CSIC-IEEC),
Campus UAB, Torre C5-Parell-2a pl, E-08193
Bellaterra (Barcelona) Spain \\
Instituci\`{o} Catalana de Recerca i Estudis Avan\c{c}ats
(ICREA), Barcelona, Spain \\
Tomsk State Pedagogical University, 634061,Tomsk, Russia \\
\email{odintsov@ieec.uab.es}}

\maketitle

\begin{history}
\received{(Day Month Year)}
\revised{(Day Month Year)}
\end{history}

\begin{abstract}

We consider modified gravity which may describe the early-time inflation
and/or late-time cosmic acceleration of the universe. In particular, we
discuss the properties of $F(R)$, $F(G)$, string-inspired and
scalar-Einstein-Gauss-Bonnet gravities, including their FRW equations and
fluid or scalar-tensor description. Simplest accelerating cosmologies are
investigated and possibility of unified description of the inflation with
dark energy is described. The cosmological reconstruction program which
permits to get the requested universe evolution from modified gravity is
developed.
As some extension, massive $F(R)$ bigravity which is ghost-free theory is
presented. Its scalar-tensor form turns out to be the easiest formulation.
The cosmological reconstruction method for such bigravity is presented.
The unified description of inflation with dark energy in $F(R)$ bigravity 
turns out to be possible.

\end{abstract}


\section{Introduction}

Modified gravity became the essential part of theoretical cosmology.
It is proposed as generalization of General Relativity with the purpose to
understand the qualitative change of gravitational interaction in the very
early and/or very late universe.
In particular, it is accepted nowadays that
modified gravity may not only describe the early-time inflation and
late-time acceleration but also may propose the unified consistent
description of the universe evolution epochs sequence: inflation,
radiation/matter dominance and dark energy. Despite the number of efforts 
to show that only General Relativity is consistent theory,
the viable and realistic models of modified gravity which pass local tests
as well as cosmological bounds are constructed.

In this contribution, we review several most popular models of alternative
gravity and discuss their cosmological applications. In the next section
we start from the most popular $F(R)$ gravity which contains higher 
derivative terms but is known to be ghost-free
theory. After short description of its elementary properties and
simplest (accelerating) cosmological solutions we introduce the
corresponding fluid and scalar-tensor description of $F(R)$ gravity.
The viability conditions of the model are outlined. Several realistic
theories which pass these viability conditions are described. In 
particular, it is shown that fifth force is not generated in these models.
Third section is devoted to modified Gauss-Bonnet gravity. Its fluid
description is introduced and field equations are presented.
For simplest, power-law model the accelerating cosmological solutions are
described. The possibility of the unification of inflation with dark
energy is briefly mentioned. The non-inducing of the correction to the Newton
law is remarkable in such a theory. Finally, the relation with 
scalar-Gauss-Bonnet gravity is explicitly outlined. 
The cosmological reconstruction program is presented in all the detail. 
Section four is devoted to the investigation
of string-inspired gravities. Scalar-Einstein-Gauss-Bonnet gravity is
considered as one of such models. For specific choice of (exponential)
scalar potential the late-time cosmic acceleration is realized in terms of
such Gauss-Bonnet dark energy. In the section five we present the version
of massive $F(R)$ bigravity which does not contain massive ghost. Its
properties are described in two representations: with and without scalars.
The appearance of two metrics (physical and reference ones) in such
formulation is presented.
The cosmological reconstruction program for such theory is
developed in detail. It is shown how one can get arbitrary accelerating
physical universe within above scheme. Finally, some summary and outlook
is given in Discussion.

\section{$F(R)$ gravity \label{IIA}}

In this section we give the elementary introduction to $F(R)$ gravity
properties and cosmology following in part the reviews
\cite{Nojiri:2006ri,Nojiri:2010wj,Capozziello:2010zz,Capozziello:2007ec}
where more complete discussion maybe found.

\subsection{General properties}

In the action of $F(R)$ gravity, the scalar curvature $R$ in
the Einstein-Hilbert action is replaced by an appropriate function
of the scalar curvature:
\be
\label{JGRG7}
S_{F(R)}= \int d^4 x \sqrt{-g} \left( \frac{F(R)}{2\kappa^2}
+ \mathcal{L}_\mathrm{matter} \right)\, .
\ee
Let us review the general properties of $F(R)$ gravity.
For $F(R)$ theory, we may define an effective EoS parameter
using its fluid representation \cite{Nojiri:2010wj}.
The FRW equations in the Einstein gravity coupled with perfect fluid
are given by
\be
\label{JGRG11}
\rho_\mathrm{matter}=\frac{3}{\kappa^2}H^2 \, ,\quad p_\mathrm{matter}
= - \frac{1}{\kappa^2}\left(3H^2 + 2\dot H\right)\, ,
\ee
which allow us to define an effective equation of state (EoS)
parameter as follows:
\be
\label{JGRG12}
w_\mathrm{eff}= - 1 - \frac{2\dot H}{3H^2} \, .
\ee

The field equation in the $F(R)$ gravity with matter is given by
\be
\label{JGRG13}
\frac{1}{2}g_{\mu\nu} F(R) - R_{\mu\nu} F'(R) - g_{\mu\nu} \Box F'(R)
+ \nabla_\mu \nabla_\nu F'(R) = - \frac{\kappa^2}{2}T_{\mathrm{matter}\,
\mu\nu}\, .
\ee
By assuming a spatially flat FRW universe,
\be
\label{JGRG14}
ds^2 = - dt^2 + a(t)^2 \sum_{i=1,2,3} \left(dx^i\right)^2\, ,
\ee
the equations corresponding to the FRW equations are given as follows:
\begin{align}
\label{JGRG15}
0 =& -\frac{F(R)}{2} + 3\left(H^2 + \dot H\right) F'(R)
  - 18 \left( 4H^2 \dot H + H \ddot H\right) F''(R)
+ \kappa^2 \rho_\mathrm{matter}\, ,\\
\label{Cr4b}
0 =& \frac{F(R)}{2} - \left(\dot H + 3H^2\right)F'(R)
+ 6 \left( 8H^2 \dot H + 4 {\dot H}^2 + 6 H \ddot H + \dddot H\right) F''(R)
\nn
&+ 36\left( 4H\dot H + \ddot H\right)^2 F'''(R) 
+ \kappa^2 p_\mathrm{matter}\, .
\end{align}
Here, the Hubble rate $H$ is defined by $H=\dot a/a$ and
the scalar curvature $R$ is given by $R=12H^2 + 6\dot H$.

One can find several (often exact) solutions of (\ref{JGRG15}).
When we neglect the contribution from matter,
by assuming that the Ricci tensor is covariantly constant,
that is, $R_{\mu\nu}\propto g_{\mu\nu}$, Eq.~(\ref{JGRG13})
reduces to an algebraic equation:
\be
\label{JGRG16}
0 = 2 F(R) - R F'(R)\, .
\ee
If Eq.~(\ref{JGRG16}) has a solution, the (anti-)de Sitter,
the Schwarzschild-(anti-)de Sitter space, and/or the Kerr-(anti-)de Sitter
space is an exact vacuum solution.

Now we assume that $F(R)$ behaves as $F(R) \propto f_0 R^m$.
Then Eq.~(\ref{JGRG15}) gives
\begin{align}
\label{M7}
0 =& f_0 \left\{ - \frac{1}{2} \left(6\dot H + 12 H^2\right)^m
+ 3 m \left(\dot H + H^2\right)\left(6\dot H + 12 H^2\right)^{m - 1} \right.
\nn
& \left. -3 m H \frac{d}{dt} \left\{\left(6\dot H + 12 H^2
\right)^{m -1}\right\}\right\} + \kappa^2 \rho_0 a^{-3(1+w)}\, .
\end{align}
Eq.~(\ref{Cr4b}) is irrelevant because it can be derived from (\ref{M7}).
When the contribution from the matter can be neglected ($\rho_0=0$),
the following solution exists:
\be
\label{JGRG17}
H \sim \frac{-\frac{(m-1)(2m-1)}{m-2}}{t}\, ,
\ee
which corresponds to the following EoS parameter (\ref{JGRG12}):
\be
\label{JGRG18}
w_\mathrm{eff}=-\frac{6m^2 - 7m - 1}{3(m-1)(2m -1)}\, .
\ee
On the other hand, when the matter with a
constant EoS parameter $w$ is included, an exact solution of (\ref{M7}) is
given by
\begin{align}
\label{M8}
& a=a_0 t^{h_0} \, ,\quad h_0\equiv \frac{2m}{3(1+w)} \, ,\nn
& a_0\equiv \left[-\frac{3f_0h_0}{\kappa^2 \rho_0}\left(-6h_0 + 12
h_0^2\right)^{m-1}
\left\{\left(1-2m\right)\left(1-m\right) -
(2-m)h_0\right\}\right]^{-\frac{1}{3(1+w)}}\, ,
\end{align}
and we find the effective EoS parameter (\ref{JGRG12}) as
\be
\label{JGRG20}
w_\mathrm{eff}= -1 + \frac{w+1}{m}\, .
\ee
These solutions (\ref{JGRG17}) and (\ref{M8}) show that modified gravity
may describe early/late-time universe acceleration.
Furthermore, it is very natural to propose that a more complicated modified
gravity from the above class may give the unified description for inflation
with late-time acceleration.

\subsubsection{Scalar-tensor description}

One can rewrite $F(R)$ gravity as the scalar-tensor theory.
By introducing the auxiliary field $A$, the action (\ref{JGRG7}) of
the $F(R)$ gravity is rewritten in the following form:
\be
\label{JGRG21}
S=\frac{1}{2\kappa^2}\int d^4 x \sqrt{-g} \left\{F'(A)\left(R-A\right)
+ F(A)\right\}\, .
\ee
By the variation of $A$, one obtains $A=R$. Substituting $A=R$ into
the action (\ref{JGRG21}), one can reproduce the action in (\ref{JGRG7}).
Furthermore, by rescaling the metric as
$g_{\mu\nu}\to \e^\sigma g_{\mu\nu}$ $\left(\sigma = -\ln F'(A)\right)$,
we obtain the Einstein frame action:
\begin{align}
\label{JGRG23}
S_E =& \frac{1}{2\kappa^2}\int d^4 x \sqrt{-g}
\left( R - \frac{3}{2}g^{\rho\sigma}
\partial_\rho \sigma \partial_\sigma \sigma - V(\sigma)\right) \, ,\nn
V(\sigma) =& \e^\sigma g\left(\e^{-\sigma}\right)
  - \e^{2\sigma} f\left(g\left(\e^{-\sigma}\right)\right)
= \frac{A}{F'(A)} - \frac{F(A)}{F'(A)^2}\, .
\end{align}
Here $g\left(\e^{-\sigma}\right)$ is given by solving the equation
$\sigma = -\ln\left( 1 + f'(A)\right)=- \ln F'(A)$ as
$A=g\left(\e^{-\sigma}\right)$.
Due to the conformal transformation, a coupling
of the scalar field $\sigma$ with usual matter arises.
What are the properties of this scalar field? Since the mass of $\sigma$ is
given by
\be
\label{JGRG24}
m_\sigma^2 \equiv \frac{3}{2}\frac{d^2 V(\sigma)}{d\sigma^2}
=\frac{3}{2}\left\{\frac{A}{F'(A)} - \frac{4F(A)}{\left(F'(A)\right)^2}
+ \frac{1}{F''(A)}\right\}\, ,
\ee
unless $m_\sigma$ is very large, the large correction to the Newton law
appears.
We can naively expect the order of the mass $m_\sigma$ to be that of the
Hubble rate, that is, $m_\sigma \sim H \sim 10^{-33}\,\mathrm{eV}$,
which is very light and could make the correction very large (fifth force
appearance).

As an example, we may consider the following exponential model
\cite{Cognola:2007zu} (see also \cite{Bamba:2010ws}
\be
\label{RDE1}
F(R) = R + \alpha \left( \e^{- bR} - 1 \right)\, .
\ee
Here $\alpha$ and $b$ are constants. One can regard $\alpha$ as an effective
cosmological constant and we choose the parameter $b$ so that
$1/b$ is much smaller than the curvature $R_0$ of the present universe.
Then in the region $R\gg R_0$, we find
\be
\label{RDE2}
m_\sigma^2 \sim \frac{\e^{bR}}{2 \alpha b^2}\, ,
\ee
which is positive and $m_\sigma^2$ could be very large and the correction
to the Newton law is very small.

In Ref.~.\cite{Hu:2007nk}, the one of the first examples of ``realistic'' $F(R)$
model was proposed.
It has been found, however, that the model has an instability where
the large curvature can be easily produced (manifestation of a
possible future singularity).
In the model of \cite{Hu:2007nk}, a parameter
$m\sim 10^{-33}\, \mathrm{eV}$ with a mass dimension is included.
The parameter $m$ plays a role of the effective cosmological constant.
When the curvature $R$ is large enough compared with $m^2$, $R\gg m^2$,
$F(R)$ in \cite{Hu:2007nk} behaves as follows:
\be
\label{HS1}
F(R) = R - c_1 m^2 + \frac{c_2 m^{2n+2}}{R^n}
+ \mathcal{O}\left(R^{-2n}\right)\, .
\ee
Here $c_1$, $c_2$, and $n$ are positive dimensionless constants.
The potential $V(\sigma)$ (\ref{JGRG23}) corresponding to (\ref{HS1})
has the following asymptotic form:
\be
\label{HS2}
V(\sigma) \sim \frac{c_1 m^2}{A^2}\, .
\ee
Then, the infinite curvature $R=A\to \infty$ corresponds to a small
value of the potential and, therefore, the large curvature can be
easily produced.

Let us assume that, when $R$ is large, $F(R)$ behaves as
\be
\label{HS3}
F(R) \sim F_0 R^{\epsilon}\, .
\ee
Here, $F_0$ and $\epsilon$ are positive constants.
We also assume $\epsilon>1$ so that this term
dominates compared with general relativity when the curvature is large.
Then, the potential $V(\sigma)$(\ref{JGRG23}) behaves as
\be
\label{HS4}
V(\sigma) \sim \frac{\epsilon -1}{\epsilon^2 F_0 R^{\epsilon -2}}\, .
\ee
Therefore, if $1<\epsilon<2$, the potential $V(\sigma)$ diverges
when $R\to \infty$ and, therefore, the large curvature is not realized so
easily.
When $\epsilon=2$, $V(\sigma)$ takes a finite value $1/F_0$ when
$R\to \infty$.
As long as $1/F_0$ is large enough, the large curvature can be prevented.

Note that the anti-gravity regime appears when $F'(R)$ is negative,
which follows from Eq.~(\ref{JGRG21}) \cite{Nojiri:2003ft}.
Then, we need to require
\be
\label{FR1}
F'(R) > 0 \, .
\ee

We should also note that
\be
\label{FRV1}
\frac{dV(\sigma)}{dA} = \frac{F''(A)}{F'(A)^3}
\left( - AF'(A) + 2F(A) \right)\, .
\ee
Therefore, if
\be
\label{FRV2}
0 = - AF'(A) + 2F(A) \, ,
\ee
the scalar field $\sigma$ is on the local maximum or local minimum of the
potential and, therefore, $\sigma$ can be a constant.
Note that the condition (\ref{FRV2}) is nothing but the
condition (\ref{JGRG16}) for the existence of the de Sitter solution.
When the condition (\ref{FRV2}) is satisfied, the mass (\ref{JGRG24}) can
be rewritten as
\be
\label{FRV3}
m_\sigma^2 = \frac{3}{2 F'(A)} \left( - A
+ \frac{F'(A)}{F''(A)} \right)\, .
\ee
Then, when the condition (\ref{FR1}) for the non-existence of the
anti-gravity is satisfied, the mass squared $m_\sigma^2$ is positive and,
therefore the, scalar field is on the local minimum if
\be
\label{FRV4}
  - A + \frac{F'(A)}{F''(A)} > 0\, .
\ee
On the other hand, if
\be
\label{FRV5}
  - A + \frac{F'(A)}{F''(A)} < 0\, ,
\ee
the scalar field is on the local maximum of the potential and
the mass squared $m_\sigma^2$ is negative.
As we will see later, the condition (\ref{FRV4}) is nothing but the
condition for stability of the de Sitter space.

Although we have rewritten the action (\ref{JGRG7}) of $F(R)$ gravity
into a scalar-tensor form (\ref{JGRG23}), inversely, it is always possible
to rewrite the action of the scalar-tensor theory
as the action of $F(R)$ gravity \cite{Capozziello:2005mj}.

Finally in this subsection, we should note that, for the transformed metric,
even if the Einstein frame universe is in a non-phantom phase,
where the effective EoS $w_\mathrm{eff}$ in (\ref{JGRG12}) is
larger than $-1$, the Jordan frame universe can be, in general, in a
phantom phase.
This apparent discrepancy occurs due to the fact that the conformal
transformation changes the interval of cosmological time.
We should note, however, the time interval which a clock measures
is not changed by the
conformal transformation.

\subsubsection{Viable modified gravities}

In order to obtain a realistic and viable model,
$F(R)$ gravity should satisfy the following conditions:
\begin{enumerate}
\item\label{req1} When $R\to 0$,
the Einstein gravity is recovered, that is,
\be
\label{E1}
F(R) \to R \quad \mbox{that is,} \quad \frac{F(R)}{R^2}
\to \frac{1}{R}\, .
\ee
This also means that there is a flat space solution.
\item\label{req2} There appears a stable de Sitter solution,
which corresponds to the late-time acceleration and, therefore, the
curvature is small
$R\sim R_L \sim \left( 10^{-33}\, \mathrm{eV}\right)^2$.
This requires, when $R\sim R_L$,
\be
\label{E2}
\frac{F(R)}{R^2} = f_{0L} - f_{1L} \left( R - R_L \right)^{2n+2}
+ o \left( \left( R - R_L \right)^{2n+2} \right)\, .
\ee
Here, $f_{0L}$ and $f_{1L}$ are positive constants and $n$ is a positive
integer. Of course, in some cases this condition may not be strictly
necessary.
\item\label{req3} There appears a quasi-stable de Sitter solution that
corresponds
to the inflation of the early universe and, therefore, the curvature is large
$R\sim R_I \sim \left( 10^{16 \sim 19}\, \mathrm{GeV}\right)^2$.
The de Sitter space should not be exactly stable so that the curvature
decreases
very slowly. It requires
\be
\label{E3}
\frac{F(R)}{R^2} = f_{0I} - f_{1I} \left( R - R_I \right)^{2m+1}
+ o \left( \left( R - R_I \right)^{2m+1} \right)\, .
\ee
Here, $f_{0I}$ and $f_{1I}$ are positive constants and $m$ is a positive
integer.
\item\label{req4} Following the discussion after (\ref{HS3}),
in order to avoid the curvature singularity when $R\to \infty$,
$F(R)$ should behaves as
\be
\label{E4}
F(R) \to f_\infty R^2 \quad \mbox{that is} \quad \frac{F(R)}{R^2} \to
f_\infty
\, .
\ee
Here, $f_\infty$ is a positive and sufficiently small constant.
Instead of (\ref{E4}), we may take
\be
\label{E5}
F(R) \to f_{\tilde \infty} R^{2 - \epsilon} \quad \mbox{that is}
\quad \frac{F(R)}{R^2} \to \frac{f_{\tilde\infty}}{R^\epsilon} \, .
\ee
Here, $f_{\tilde\infty}$ is a positive constant and $0< \epsilon <1$.
The above condition (\ref{E4}) or (\ref{E5}) prevents both the future
singularity \cite{Bamba:2008ut} and the singularity due to large density
of matter.
\item\label{req5} As in (\ref{FR1}), to avoid the anti-gravity,
we require
\be
\label{E6}
F'(R)>0\, ,
\ee
which is rewritten as
\be
\label{E7}
\frac{d}{dR} \left( \ln \left( \frac{F(R)}{R^2} \right)\right)
> - \frac{2}{R}\, .
\ee
\item\label{req6}
Combining conditions (\ref{E1}) and (\ref{E6}), one finds
\be
\label{E8}
F(R)>0\, .
\ee
\item To avoid the matter instability \cite{Dolgov:2003px},
we require
\begin{align}
\label{E8B}
U(R_b) \equiv& \frac{R_b}{3} - \frac{F^{(1)}(R_b) F^{(3)}(R_b) R_b}
{3 F^{(2)}(R_b)^2} - \frac{F^{(1)}(R_b)}{3F^{(2)}(R_b)} \nn
& + \frac{2 F(R_b) F^{(3)}(R_b)}{3 F^{(2)}(R_b)^2} - \frac{F^{(3)}(R_b) R_b}{3
F^{(2)}(R_b)^2}< 0\, .
\end{align}
\end{enumerate}
The conditions \ref{req1} and \ref{req2} tell that an extra, unstable de Sitter
solution must appear at $R=R_e$ $\left( 0< R_e < R_L \right)$.
Since the universe evolution will stop at $R=R_L$ because the de Sitter
solution $R=R_L$ is stable; the curvature never becomes smaller than $R_L$
and, therefore, the extra de Sitter solution is not realized.

An example of viable $F(R)$ gravity is given in \cite{Nojiri:2010ny}
\begin{align}
\label{EE1}
\frac{F(R)}{R^2} =&
\left\{ \left(X_m \left(R_I;R\right) - X_m\left(R_I;R_1\right) \right)
\left(X_m\left(R_I;R\right) - X_m\left(R_I;R_L\right) \right)^{2n+2}
\right. \nn
& \left. + X_m\left(R_I;R_1\right) X_m\left(R_I;R_L\right)^{2n+2}
+ f_\infty^{2n+3} \right\}^{\frac{1}{2n+3}} \, , \nn
X_m\left(R_I;R\right) \equiv &
\frac{\left(2m +1\right) R_I^{2m}}{\left( R - R_I \right)^{2m+1}
+ R_I^{2m+1}} \, .
\end{align}
Here, $n$ and $m$ are integers greater or equal to unity, and $n,m\geq 1$
and $R_1$ is a parameter related with $R_e$ by
\be
\label{EE2}
X\left(R_I;R_e\right) = \frac{\left(2n+2\right)
X\left(R_I;R_1\right)X\left(R_I;R_1\right) + X\left(R_I;R_L\right)}{2n+3}\, .
\ee
We also assume $0<R_1<R_L \ll R_I$.

Another realistic theory unifying inflation with dark energy is given in
\cite{Elizalde:2010ts}
\be
\label{RDE5}
F(R) = R - 2 \Lambda \left( 1 - \e^{-\frac{R}{R_0}} \right)
  - \Lambda_i \left( 1 - \e^{ - \left( \frac{R}{R_i} \right)^n } \right)
+ \gamma R^\alpha\, .
\ee
Here $\Lambda$ is the effective cosmological constant in the present universe
and we also assume the parameter $R_0$ is almost equal to $\Lambda$.
$R_i$ and $\Lambda_i$ are typical values of the curvature and the effective
cosmological constant. $\alpha$ is a constant: $1<\alpha \leq 2$.

In the same way as above one can construct a number of viable $F(R)$ gravity
models.
These models may
explain the early-time inflation in addition to the dark energy epoch
in a unified way.

\section{$f(\mathcal{G})$ gravity \label{IIB}}

\subsubsection{General properties}

We also proposed another class of modified gravity where the arbitrary
function, which depends on topological Gauss-Bonnet invariant:
\be
\label{GB}
\mathcal{G}=R^2 -4 R_{\mu\nu} R^{\mu\nu} + R_{\mu\nu\xi\sigma}
R^{\mu\nu\xi\sigma}\, ,
\ee
is added to the action of the Einstein gravity.
One starts with the following action
\cite{Nojiri:2005jg,Nojiri:2005am,Cognola:2006eg}:
\be
\label{GB1b}
S=\int d^4x\sqrt{-g} \left(\frac{1}{2\kappa^2}R + f(\mathcal{G})
+ \mathcal{L}_\mathrm{matter}\right)\, .
\ee
Here, $\mathcal{L}_\mathrm{matter}$ is the Lagrangian density of matter.
The variation of the metric $g_{\mu\nu}$ gives an equation corresponding
to the Einstein equation:
\begin{align}
\label{GB4b} & 0= \frac{1}{2\kappa^2}\left(- R^{\mu\nu}
+ \frac{1}{2} g^{\mu\nu} R\right) + T_\mathrm{matter}^{\mu\nu}
+ \frac{1}{2}g^{\mu\nu} f(\mathcal{G}) -2 f'(\mathcal{G}) R R^{\mu\nu} \nn
& + 4f'(\mathcal{G})R^\mu_{\ \rho}
R^{\nu\rho} -2 f'(\mathcal{G}) R^{\mu\rho\sigma\tau}
R^\nu_{\ \rho\sigma\tau} - 4 f'(\mathcal{G})
R^{\mu\rho\sigma\nu}R_{\rho\sigma}
+ 2 \left( \nabla^\mu \nabla^\nu f'(\mathcal{G})\right)R \nn
& - 2 g^{\mu\nu} \left( \nabla^2 f'(\mathcal{G})\right)
R - 4 \left( \nabla_\rho \nabla^\mu f'(\mathcal{G})\right)
R^{\nu\rho} - 4 \left( \nabla_\rho \nabla^\nu
f'(\mathcal{G})\right)R^{\mu\rho} \nn
& + 4 \left( \nabla^2 f'(\mathcal{G}) \right)R^{\mu\nu} + 4g^{\mu\nu}
\left( \nabla_{\rho} \nabla_\sigma f'(\mathcal{G}) \right)
R^{\rho\sigma} - 4 \left(\nabla_\rho \nabla_\sigma f'(\mathcal{G}) \right)
R^{\mu\rho\nu\sigma} \, .
\end{align}
We should note that this equation does not contain the
terms which contain derivatives higher than of second order.

By choosing the spatially flat FRW universe metric (\ref{JGRG14}),
we obtain the equation corresponding to the first FRW equation:
\be
\label{GB7b}
0=-\frac{3}{\kappa^2}H^2 - f(\mathcal{G})
+ \mathcal{G}f'(\mathcal{G}) - 24 \dot{\mathcal{G}}f''(\mathcal{G}) H^3
+ \rho_\mathrm{matter}\, .
\ee
In the FRW universe (\ref{JGRG14}), $\mathcal{G}$ has the following form:
\be
\label{mGB000}
\mathcal{G} = 24 \left( H^2 \dot H + H^4 \right)\, .
\ee
Then, from Eq.~(\ref{GB7b}), as in the Einstein gravity case (\ref{JGRG11}),
we find the FRW-like equations (fluid description):
\be
\label{mGB1BB}
\rho^\mathcal{G}_\mathrm{eff}=\frac{3}{\kappa^2}H^2 \, ,
\quad p^\mathcal{G}_\mathrm{eff}= - \frac{1}{\kappa^2}\left(3H^2
+ 2\dot H\right)\, .
\ee
Here,
\begin{align}
\label{mGB2BB}
\rho^\mathcal{G}_\mathrm{eff} \equiv& - f(\mathcal{G}) +
\mathcal{G}f'(\mathcal{G}) - 24 \dot{\mathcal{G}}f''(\mathcal{G}) H^3
+ \rho_\mathrm{matter}\, , \nn
p^\mathcal{G}_\mathrm{eff} \equiv&
f(\mathcal{G}) - \mathcal{G}f'(\mathcal{G})
+ \frac{2 \mathcal{G} \dot{\mathcal{G}}}{3H} f''(\mathcal{G})
+ 8 H^2 \ddot{\mathcal{G}} f''(\mathcal{G})
+ 8 H^2 {\dot{\mathcal{G}}}^2 f'''(\mathcal{G}) + p_\mathrm{matter}\, .
\end{align}
When $\rho_\mathrm{matter}=0$, Eq.~(\ref{GB7b}) has a de Sitter universe
solution where $H$, and therefore $\mathcal{G}$, are constant. For
$H=H_0$, with a constant $H_0$, Eq.~(\ref{GB7b}) turns into
\be
\label{GB7bb}
0=-\frac{3}{\kappa^2}H_0^2 + 24H_0^4
f'\left( 24H_0^4 \right) - f\left( 24H_0^4\right) \, .
\ee
As an example, we consider the model
\be
\label{mGB1}
f(\mathcal{G})=f_0\left|\mathcal{G}\right|^\beta\, ,
\ee
with constants $f_0$ and $\beta$. Then, the solution of Eq.~(\ref{GB7bb}) is
given by
\be
\label{GBGB2}
H_0^4 = \frac{1}{24 \left( 8 \left(n-1\right) \kappa^2 f_0
\right)^{\frac{1}{\beta - 1}}}\, .
\ee
For a large number of choices of the function $f(\mathcal{G})$,
Eq.~(\ref{GB7bb}) has a non-trivial ($H_0\neq 0$) real solution
for $H_0$ (de Sitter universe).
The late-time cosmology for above theory without matter has been first
discussed for a number of examples
in Refs.~\cite{Nojiri:2005jg,Nojiri:2005am,Cognola:2006eg}.

Now, we consider the case in which the contributions from the Einstein and
matter
terms can be neglected. Eq.~(\ref{GB7b}) reduces to
\be
\label{mGB9}
0=\mathcal{G}f'(\mathcal{G}) - f(\mathcal{G}) - 24 \dot{\mathcal{G}}
f''(\mathcal{G}) H^3 \, .
\ee
If $f(\mathcal{G})$ behaves as (\ref{mGB1}), assuming
\be
\label{mGB2}
a=\left\{\begin{array}{ll} a_0t^{h_0}\ &\mbox{when}\ h_0>0\
\mbox{(quintessence)} \\
a_0\left(t_s - t\right)^{h_0}\ &\mbox{when}\ h_0<0\ \mbox{(phantom)} \\
\end{array} \right. \, ,
\ee
one obtains
\be
\label{mGB10}
0=\left(\beta - 1\right)h_0^6\left(h_0 - 1\right)
\left(h_0 - 1 + 4\beta \right)\, .
\ee
As $h_0=1$ implies $\mathcal{G}=0$, one may choose
\be
\label{mGB11}
h_0 = 1 - 4\beta\, ,
\ee
and Eq.~(\ref{JGRG12}) gives
\be
\label{mGB12}
w_\mathrm{eff}=-1 + \frac{2}{3(1-4\beta)}\, .
\ee
Therefore, if $\beta>0$, the universe is accelerating
($w_\mathrm{eff}<-1/3$), and if $\beta>1/4$, the universe is in
a phantom phase ($w_\mathrm{eff}<-1$).
Thus, we are led to consider the following model:
\be
\label{mGB13}
f(\mathcal{G})=f_i\left|\mathcal{G}\right|^{\beta_i}
+ f_l\left|\mathcal{G}\right|^{\beta_l} \, ,
\ee
where it is assumed that
\be
\label{mGB14}
\beta_i>\frac{1}{2}\, ,\quad \frac{1}{2}>\beta_l>\frac{1}{4}\, .
\ee
Then, when the curvature is large, as in the primordial universe, the first
term dominates, compared with the second term and the Einstein term,
and it gives
\be
\label{mGB15}
  -1>w_\mathrm{eff}=-1
+ \frac{2}{3(1-4\beta_i)}>- \frac{5}{3}\, .
\ee
On the other hand, when the curvature is small, as is the case in
the present universe, the second term in (\ref{mGB13}) dominates
compared with the first term and the Einstein term and yields
\be
\label{mGB16} w_\mathrm{eff}= -1
+ \frac{2}{3(1-4\beta_l)}< - \frac{5}{3}\, .
\ee
Therefore, theory (\ref{mGB13}) can produce a model that is
able to describe inflation and the late-time acceleration
of the universe in a unified manner.

Instead of (\ref{mGB14}), one may also choose $\beta_l$ as
\be
\label{mGB17}
\frac{1}{4}>\beta_l>0\, ,
\ee
which gives
\be
\label{mGB18} -\frac{1}{3}>w_\mathrm{eff}>-1\, .
\ee
Then, we obtain an effective quintessence epoch. Moreover, by
properly adjusting the couplings $f_i$ and $f_l$ in (\ref{mGB13}),
one can obtain a period where the Einstein term dominates and
the universe is in a deceleration phase. After that, a transition
occurs from deceleration to acceleration when the Gauss-Bonnet term
becomes the dominant one.
More choices of $f(\mathcal{G})$ may be studied for the purpose of
the construction of the current accelerating universe. Nevertheless, many
non-linear choices for this function may be approximated by the above model.
For instance, one can mention some realistic examples of $f(\mathcal{G})$
gravity:
\be
f_{1}(\mathcal{G}) =
\frac{a_{1}\mathcal{G}^{n}+b_{1}}{a_{2}\mathcal{G}^{n}+b_{2}}\, ,
\quad f_{2}(\mathcal{G}) =
\frac{a_{1}\mathcal{G}^{n+N}+b_{1}}{a_{2}\mathcal{G}^{n}+b_{2}}\,.
\label{due}
\ee

We now address the issue of the correction to the Newton law. Let
$g_{(0)}$ be a solution of (\ref{GB4b}) and represent the
perturbation of the metric as
$g_{\mu\nu}=g_{(0)\mu\nu} + h_{\mu\nu}$.
First, we consider the perturbation around the de Sitter
background. The de Sitter space metric is taken as $g_{(0)\mu\nu}$,
which gives the following Riemann tensor:
\be
\label{GB35}
R_{(0)\mu\nu\rho\sigma}=H_0^2\left(g_{(0)\mu\rho}
g_{(0)\nu\sigma} - g_{(0)\mu\sigma}g_{(0)\nu\rho}\right)\, .
\ee
The flat background corresponds to the limit of $H_0\to 0$. For
simplicity, the following gauge condition is chosen:
$g_{(0)}^{\mu\nu} h_{\mu\nu}=\nabla_{(0)}^\mu h_{\mu\nu}=0$. Then
Eq.~(\ref{GB4b}) gives
\be
\label{GB38b}
0=\frac{1}{4\kappa^2} \left(\nabla^2 h_{\mu\nu} - 2H_0^2 h_{\mu\nu}\right)
+ T_{\mathrm{matter}\, \mu\nu}\, .
\ee
The Gauss-Bonnet term contribution does not appear except
in the length parameter $1/H_0$ of the de Sitter space, which is
determined by taking into account the Gauss-Bonnet term.
This may occur due to the special structure of the Gauss-Bonnet invariant.
Eq.~(\ref{GB38b}) shows that there is no correction to the Newton law in
de Sitter space and even in the flat background corresponding to
$H_0\to 0$, regardless of the form of $f$ (at least, with the above
gauge condition).

The action (\ref{GB1b}) can be rewritten by introducing
the auxiliary scalar field $\phi$ as \cite{Nojiri:2006je,Cognola:2006sp},
\be
\label{fG2}
S=\int d^4 x \sqrt{-g}\left[ \frac{R}{2\kappa^2} - V(\phi)
  - \xi(\phi) \mathcal{G} \right]\, .
\ee
By variation over $\phi$, one obtains
\be
\label{fG3}
0=V'(\phi) + \xi'(\phi) \mathcal{G}\, ,
\ee
which could be solved with respect to $\phi$ as
\be
\label{fG4}
\phi= \phi(\mathcal{G})\, .
\ee
By substituting the expression (\ref{fG4}) into the action
(\ref{fG2}), we obtain the action of $f(\mathcal{G})$ gravity, with
\be
\label{fG5}
f(\mathcal{G})= - V\left(\phi(\mathcal{G})\right)
+ \xi\left(\phi(\mathcal{G})\right)\mathcal{G}\, .
\ee
Assuming a spatially-flat FRW universe and the scalar field $\phi$
to depend only on $t$, we obtain the field equations:
\begin{align}
\label{fG6}
0=& - \frac{3}{\kappa^2}H^2 + V(\phi) + 24 H^3 \frac{d \xi(\phi(t))}{dt}\, ,\\
\label{fG7}
0=& \frac{1}{\kappa^2}\left(2\dot H + 3 H^2 \right) - V(\phi)
  - 8H^2 \frac{d^2 \xi(\phi(t))}{dt^2} \nn
& - 16H \dot H
\frac{d\xi(\phi(t))}{dt} - 16 H^3 \frac{d \xi(\phi(t))}{dt}\, .
\end{align}
Combining the above equations, we obtain
\begin{align}
\label{fG8}
0=& \frac{2}{\kappa^2}\dot H - 8H^2 \frac{d^2
\xi(\phi(t))}{dt^2} - 16 H\dot H \frac{d\xi(\phi(t))}{dt}
+ 8H^3 \frac{d\xi(\phi(t))}{dt} \nn
=& \frac{2}{\kappa^2}\dot H - 8a\frac{d}{dt}\left(
\frac{H^2}{a}\frac{d\xi(\phi(t))}{dt}\right)\, ,
\end{align}
which can be solved with respect to $\xi(\phi(t))$ as
\be
\label{fG9}
\xi(\phi(t))=\frac{1}{8}\int^t dt_1 \frac{a(t_1)}{H(t_1)^2} W(t_1)\, ,\quad
W(t)\equiv \frac{2}{\kappa^2} \int^{t} \frac{dt_1}{a(t_1)} \dot H (t_1)\, .
\ee
Combining (\ref{fG6}) and (\ref{fG9}), the expression
for $V(\phi(t))$ follows:
\be
\label{fG10}
V(\phi(t)) = \frac{3}{\kappa^2}H(t)^2 - 3a(t) H(t) W(t)\, .
\ee
As there is a freedom of redefinition of the scalar field $\phi$,
we may identify $t$ with $\phi$.
Hence, we consider the model where $V(\phi)$ and $\xi(\phi)$ can be
expressed in terms of a single function $g$ as
\begin{align}
\label{fG11}
V(\phi) =& \frac{3}{\kappa^2}g'\left(\phi\right)^2 - 3g'\left(\phi\right)
\e^{g\left(\phi\right)} U(\phi) \, , \nn
\xi(\phi) =& \frac{1}{8}\int^\phi d\phi_1
\frac{\e^{g\left(\phi_1\right)} }{g'(\phi_1)^2} U(\phi_1)\, ,\nn
U(\phi) \equiv& \frac{2}{\kappa^2}\int^\phi d\phi_1
\e^{-g\left(\phi_1\right)} g''\left(\phi_1\right) \, .
\end{align}
By choosing $V(\phi)$ and $\xi(\phi)$ as (\ref{fG11}), one can easily
find
the following solution for Eqs.(\ref{fG6}) and (\ref{fG7}):
\be
\label{fGB12}
a=a_0\e^{g(t)}\ \left(H= g'(t)\right)\, .
\ee
Therefore one can reconstruct $F(G)$ gravity to generate
arbitrary expansion history of the universe.

Thus, we reviewed the modified Gauss-Bonnet gravity and demonstrated that
it may naturally lead to the unified cosmic history, including the inflation
and dark energy era.

\section{String-inspired model and scalar-Einstein-Gauss-Bonnet
gravity \label{IIC}}

In string theories, the compactification from higher dimensions to
four dimensions induces many scalar fields, such as moduli or
dilaton fields. These scalars couple with curvature invariants.
Neglecting the moduli fields associated with the radii of the internal
space, we may consider the following action of the low-energy
effective string theories \cite{Sami:2005zc,Calcagni:2005im}:
\be
\label{eq:action}
S = \int d^4 x
\sqrt{-g}\left[\frac{R}{2}+\mathcal{L}_{\phi}
+ \mathcal{L}_{c}+\ldots\right]\, ,
\ee
where $\phi$ is the dilaton field, which is related to the
string coupling, $\mathcal{L}_{\phi}$ is the Lagrangian of $\phi$,
and $\mathcal{L}_c$ expresses the string curvature correction terms
to the Einstein-Hilbert action,
\be
\label{stcor}
\mathcal{L}_{\phi} = -\partial_{\mu}\phi\partial^{\mu}\phi-V(\phi) \, ,
\quad
\mathcal{L}_c = c_1 {\alpha'} \e^{2\frac{\phi}{\phi_0}}\mathcal{L}_c^{(1)}
+ c_2{\alpha'}^2\e^{4\frac{\phi}{\phi_0}}\mathcal{L}_c^{(2)}
+ c_3{\alpha'}^3\e^{6\frac{\phi}{\phi_0}}\mathcal{L}_c^{(3)}\, ,
\ee
where $1/\alpha'$ is the string tension, $\mathcal{L}_c^{(1)}$,
$\mathcal{L}_c^{(2)}$, and $\mathcal{L}_c^{(3)}$ express the
leading-order (Gauss-Bonnet term $\mathcal{G}$ in (\ref{GB})),
the second-order, and the third-order curvature corrections, respectively.
The terms $\mathcal{L}_c^{(1)}$, $\mathcal{L}_c^{(2)}$ and
$\mathcal{L}_c^{(3)}$ in the Lagrangian have the following form
\be
\label{ccc}
\mathcal{L}_c^{(1)} = \Omega_2\, , \quad
\mathcal{L}_c^{(2)} = 2 \Omega_3 + R^{\mu\nu}_{\alpha \beta}
R^{\alpha\beta}_{\lambda\rho}
R^{\lambda\rho}_{\mu\nu}\, , \quad
\mathcal{L}_c^{(3)}
= \mathcal{L}_{31} - \delta_{H} \mathcal{L}_{32} -\frac{\delta_{B}}{2}
\mathcal{L}_{33}\, .
\ee
Here, $\delta_B$ and $\delta_H$ take the value of $0$ or $1$ and
\bea
\Omega_2 &=& \mathcal{G} \, , \nn
\Omega_3 &\propto & \epsilon^{\mu\nu\rho\sigma\tau\eta}
\epsilon_{\mu'\nu'\rho'\sigma'\tau'\eta'}
R_{\mu\nu}^{\ \ \mu'\nu'} R_{\rho\sigma}^{\ \ \rho'\sigma'}
R_{\tau\eta}^{\ \ \tau'\eta'} \, , \nn
\mathcal{L}_{31} &=& \zeta(3) R_{\mu\nu\rho\sigma}R^{\alpha\nu\rho\beta}
\left( R^{\mu\gamma}_{\ \ \delta\beta}
R_{\alpha\gamma}^{\ \ \delta\sigma} - 2 R^{\mu\gamma}_{\ \ \delta\alpha}
R_{\beta\gamma}^{\ \ \delta\sigma} \right)\, , \nn
\mathcal{L}_{32} &=& \frac{1}{8} \left( R_{\mu\nu\alpha\beta}
R^{\mu\nu\alpha\beta}\right)^2
+ \frac{1}{4} R_{\mu\nu}^{\ \ \gamma\delta}
R_{\gamma\delta}^{\ \ \rho\sigma} R_{\rho\sigma}^{\ \ \alpha\beta}
R_{\alpha\beta}^{\ \ \mu\nu} - \frac{1}{2} R_{\mu\nu}^{\ \ \alpha\beta}
R_{\alpha\beta}^{\ \ \rho\sigma}
R^\mu_{\ \sigma\gamma\delta}
R_\rho^{\ \nu\gamma\delta} - R_{\mu\nu}^{\ \ \alpha\beta}
R_{\alpha\beta}^{\ \ \rho\nu} R_{\rho\sigma}^{\ \ \gamma\delta}
R_{\gamma\delta}^{\ \ \mu\sigma}\, , \nn
\mathcal{L}_{33} &=& \left( R_{\mu\nu\alpha\beta}
R^{\mu\nu\alpha\beta}\right)^2 - 10 R_{\mu\nu\alpha\beta}
R^{\mu\nu\alpha\sigma}
R_{\sigma\gamma\delta\rho}
R^{\beta\gamma\delta\rho} - R_{\mu\nu\alpha\beta}
R^{\mu\nu\rho}_{\ \ \ \sigma}
R^{\beta\sigma\gamma\delta}
R_{\delta\gamma\rho}^{\ \ \ \alpha} \, .
\eea
The correction terms are different depending on the type of string theory;
the dependence is encoded in the curvature invariants and in
the coefficients $(c_1,c_2,c_3)$ and $\delta_H$, $\delta_B$, as follows,
\begin{itemize}
\item For the Type II superstring theory: $(c_1,c_2,c_3) = (0,0,1/8)$
and $\delta_H=\delta_B=0$.
\item For the heterotic superstring theory: $(c_1,c_2,c_3) = (1/8,0,1/8)$
and $\delta_H=1,\delta_B=0$.
\item For the bosonic superstring theory: $(c_1,c_2,c_3) = (1/4,1/48,1/8)$
and $\delta_H=0,\delta_B=1$.
\end{itemize}

Motivated by the string considerations, we consider the
scalar-Einstein-Gauss-Bonnet gravity\footnote{For pioneering work
on the scalar-Einstein-Gauss-Bonnet gravity, see \cite{Boulware:1986dr}. }
based on \cite{Nojiri:2005vv,Nojiri:2006je}. It was first proposed in
Ref.~\cite{Nojiri:2005vv} to consider such a theory as a gravitational
alternative for dark energy, so-called Gauss-Bonnet dark energy.

The starting action is:
\be
\label{ma22}
S=\int d^4 x \sqrt{-g}\left[ \frac{R}{2\kappa^2} - \frac{1}{2}
\partial_\mu \phi
\partial^\mu \phi - V(\phi) - \xi(\phi) \mathcal{G}\right]\, .
\ee
Here, we do not restrict the forms of $V(\phi)$ and $\xi(\phi)$ which
should be given by the non-perturbative string theory (\ref{stcor}).
Note also that the action (\ref{ma22}) is given by adding the kinetic
term for the scalar field $\phi$ to the action of the
$F(\mathcal{G})$ gravity in the scalar-tensor form that appeared
in the previous section.

By the variation of the action (\ref{ma22}) with respect to the metric
$g_{\mu\nu}$, we obtain the the following equations:
\begin{align}
\label{ma23}
& 0= \frac{1}{\kappa^2}\left(- R^{\mu\nu}
+ \frac{1}{2} g^{\mu\nu} R\right)
+ \frac{1}{2}\partial^\mu \phi \partial^\nu \phi - \frac{1}{4}g^{\mu\nu}
\partial_\rho \phi \partial^\rho \phi
+ \frac{1}{2}g^{\mu\nu}\left( - V(\phi) + \xi(\phi) \mathcal{G} \right)
\nn
& -2 \xi(\phi) R R^{\mu\nu} - 4\xi(\phi)R^\mu_{\ \rho}
R^{\nu\rho} -2 \xi(\phi) R^{\mu\rho\sigma\tau}R^\nu_{\ \rho\sigma\tau}
+4 \xi(\phi) R^{\mu\rho\nu\sigma}R_{\rho\sigma} \nn
& + 2 \left( \nabla^\mu \nabla^\nu \xi(\phi)\right)R - 2 g^{\mu\nu}
\left( \nabla^2\xi(\phi)\right)R - 4 \left(
\nabla_\rho \nabla^\mu \xi(\phi)\right)R^{\nu\rho} - 4 \left(
\nabla_\rho \nabla^\nu \xi(\phi)\right)R^{\mu\rho} \nn
& + 4 \left( \nabla^2 \xi(\phi) \right)R^{\mu\nu}
+ 4g^{\mu\nu} \left( \nabla_{\rho} \nabla_\sigma \xi(\phi) \right)
R^{\rho\sigma}
+ 4 \left(\nabla_\rho \nabla_\sigma \xi(\phi) \right) R^{\mu\rho\nu\sigma}
\, .
\end{align}
In Eq.~(\ref{ma23}), the derivatives of curvature such as $\nabla R$,
do not appear. Therefore, the derivatives higher than two do not appear,
which can be contrasted with a general
$\alpha R^2 + \beta R_{\mu\nu}R^{\mu\nu} + \gamma
R_{\mu\nu\rho\sigma}R^{\mu\nu\rho\sigma}$ gravity,
where fourth derivatives of $g_{\mu\nu}$ appear.
For the classical theory, if we specify the values of
$g_{\mu\nu}$ and $\dot g_{\mu\nu}$
on a spatial surface as an initial condition, the time
development is uniquely determined.
This situation is similar to the case in classical mechanics,
in which one only needs to specify the values of
position and velocity of particle as initial conditions.
In general $\alpha R^2 + \beta R_{\mu\nu}R^{\mu\nu} + \gamma
R_{\mu\nu\rho\sigma}R^{\mu\nu\rho\sigma}$ gravity,
we need to specify the values of $\ddot g_{\mu\nu}$ and $\dddot g_{\mu\nu}$
in addition to $g_{\mu\nu}$, $\dot g_{\mu\nu}$
so that a unique time development will follow.
In Einstein gravity, only the specific value of $g_{\mu\nu}$,
$\dot g_{\mu\nu}$, should be given as an initial condition.
Thus, the scalar-Gauss-Bonnet gravity is a natural extension of the
Einstein gravity.

In the FRW universe (\ref{JGRG14}), Eq.~(\ref{ma23}) becomes
the following:
\begin{align}
\label{ma24}
0=& - \frac{3}{\kappa^2}H^2 + \frac{1}{2}{\dot\phi}^2 + V(\phi)
+ 24 H^3 \frac{d \xi(\phi(t))}{dt}\, ,\\
\label{GBany5}
0=& \frac{1}{\kappa^2}\left(2\dot H + 3 H^2 \right)
+ \frac{1}{2}{\dot\phi}^2 - V(\phi) - 8H^2 \frac{d^2 \xi(\phi(t))}
{dt^2} \nn
& - 16H \dot H \frac{d\xi(\phi(t))}{dt} - 16 H^3
\frac{d \xi(\phi(t))}{dt} \, .
\end{align}
On the other hand, by the variation of the action (\ref{ma22})
with respect to the scalar field, the scalar equation
of motion follows as
\be
\label{ma24b}
0=\ddot \phi + 3H\dot \phi + V'(\phi) + \xi'(\phi) \mathcal{G}\, .
\ee

In particular when we consider the following string-inspired model
\cite{Nojiri:2005vv},
\be
\label{NOS1}
V=V_0\e^{-\frac{2\phi}{\phi_0}}\, , \quad \xi(\phi)=\xi_0
\e^{\frac{2\phi}{\phi_0}}\, ,
\ee
the de Sitter space solution follows:
\be
\label{NOS2}
H^2 = H_0^2 \equiv - \frac{\e^{-\frac{2\varphi_0}{\phi_0}}}
{8\xi_0 \kappa^2} \, , \quad \phi = \varphi_0 \, .
\ee
Here, $\varphi_0$ is an arbitrary constant.
If $\varphi_0$ is chosen to be larger, the Hubble rate $H=H_0$ becomes
smaller.
Then, if $\xi_0\sim \mathcal{O}(1)$, by choosing
$\varphi_0/\phi_0\sim 140$,
the value of the Hubble rate $H=H_0$ is consistent with the observations.
The model (\ref{NOS1}) also has another solution:
\bea
\label{NOS3}
& H=\frac{h_0}{t}\, ,\quad \phi=\phi_0 \ln \frac{t}{t_1}\
& \mbox{when}\ h_0>0\, ,\nn
& H=-\frac{h_0}{t_s - t}\, ,\quad \phi=\phi_0 \ln \frac{t_s - t}{t_1}\ &
\mbox{when}\ h_0<0\, .
\eea
Here, $h_0$ is obtained by solving the following algebraic equations:
\be
\label{NOS4}
0 = -\frac{3h_0^2}{\kappa^2} + \frac{\phi_0^2}{2}
+ V_0 t_1^2 - \frac{48 \xi_0 h_0^3}{t_1^2}\, ,\quad
0 = \left( 1 - 3h_0 \right)\phi_0^2 + 2V_0 t_1^2
+ \frac{48 \xi_0 h_0^3}{t_1^2}\left(h_0 - 1\right)\, .
\ee
Eqs.~(\ref{NOS4}) can be rewritten as
\begin{align}
\label{NOS5}
V_0 t_1^2 =& - \frac{1}{\kappa^2\left(1 + h_0\right)}\left\{3h_0^2 \left(
1 - h_0\right)
+ \frac{\phi_0^2 \kappa^2 \left( 1 - 5 h_0\right)}{2}\right\}\, ,\\
\label{NOS6}
\frac{48 \xi_0 h_0^2}{t_1^2} =& - \frac{6}{\kappa^2\left( 1
+ h_0\right)}\left(h_0 - \frac{\phi_0^2 \kappa^2}{2}\right)\, .
\end{align}
The arbitrary value of $h_0$ can be realized by properly
choosing $V_0$ and $\xi_0$.
With the appropriate choice of $V_0$ and $\xi_0$, we can obtain
a negative $h_0$ and, therefore, the effective EoS parameter
(\ref{JGRG12}) is less than $-1$, $w_\mathrm{eff} < -1$,
which corresponds to the effective phantom.
In usual (canonical) scalar-tensor theory without the Gauss-Bonnet term,
the phantom cannot be realized by the canonical scalar.

For example, if $h_0=-80/3<-1$ and, therefore, $w= - 1.025$, which is
consistent with the observed value, we find
\begin{align}
\label{NOS6b}
V_0t_1^2 =& \frac{1}{\kappa^2}\left( \frac{531200}{231}
+ \frac{403}{154}\gamma \phi_0^2 \kappa^2 \right)>0\, , \nn
\frac{f_0}{t_1^2} =& -\frac{1}{\kappa^2}\left( \frac{9}{49280}
+ \frac{27}{7884800}\gamma \phi_0^2 \kappa^2 \right)\, .
\end{align}
For other choices of scalar potentials one can realize other types of dark
energy universes, for instance, the effective quintessence. Moreover, one
can propose the potentials in such a way that the unification of
the inflation with dark energy naturally occurs. Of course, many more models of
above Gauss-Bonnet dark energy were considered. The corresponding
discussion/references maybe found in \cite{Nojiri:2010wj}.

\section{$F(R)$ bigravity}

Recently non-linear massive gravity \cite{deRham:2010ik,Hassan:2011hr} (with
non-dynamical background metric)
was extended to the ghost-free construction
with the dynamical metric \cite{Hassan:2011zd}.
The most general proof of absence of ghost in massive gravity has been given in
\cite{Hassan:2011tf}.
Especially in case of the minimal model, which we consider below in
(\ref{bimetric2}). It was first discussed in \cite{Hassan:2011vm}.
The convenient description of the theory gives bigravity or bimetric gravity
which contains two metrics (symmetric tensor fields).
One of two metrics is called physical metric while second metric is called
reference metric.

In Ref.~\cite{Nojiri:2012zu,Nojiri:2012re} we have proposed $F(R)$ bigravity
which is also ghost-free theory.
We introduce four kinds of metrics, $g_{\mu\nu}$, $g^\mathrm{J}_{\mu\nu}$,
$f_{\mu\nu}$, and $f^\mathrm{J}_{\mu\nu}$.
The physical observable metric $g^\mathrm{J}_{\mu\nu}$ is the metric in the
Jordan frame.
The metric $g_{\mu\nu}$ corresponds to the metric in the Einstein frame in
the standard $F(R)$ gravity and therefore the metric $g_{\mu\nu}$ is not
physical metric.
In the bigravity theories, we have to introduce another reference metrics or
symmetric tensor $f_{\mu\nu}$ and $f^\mathrm{J}_{\mu\nu}$.
The metric $f_{\mu\nu}$ is the metric corresponding to the Einstein frame
with respect to the curvature given by the metric $f_{\mu\nu}$.
On the other hand, the metric $f^\mathrm{J}_{\mu\nu}$ is the
metric corresponding to the Jordan frame.

\subsection{Construction of $F(R)$ bigravity \label{SII}}

In this section, we review the construction of ghost-free $F(R)$ bigravity,
following Ref.~\cite{Nojiri:2012zu}.
The consistent model of bimetric gravity, which includes two metric tensors
$g_{\mu\nu}$ and $f_{\mu\nu}$, was proposed in Ref.~\cite{Hassan:2011zd}.
It contains the massless spin-two field, corresponding to graviton, and
massive spin-two field.
The gravity model which only contains the massive spin-two field is called
massive gravity. We consider the model including both of massless and
massive spin two field,i.e. bigravity.
It has been shown that the Boulware-Deser ghost \cite{Boulware:1974sr} does not
appear in such a theory.

The starting action is given by
\begin{align}
\label{bimetric}
S_\mathrm{bi} =&M_g^2\int d^4x\sqrt{-\det g}\,R^{(g)}+M_f^2\int d^4x
\sqrt{-\det f}\,R^{(f)} \nonumber \\
&+2m^2 M_\mathrm{eff}^2 \int d^4x\sqrt{-\det g}\sum_{n=0}^{4} \beta_n\,
e_n \left(\sqrt{g^{-1} f} \right) \, .
\end{align}
Here $R^{(g)}$ is the scalar curvature for $g_{\mu \nu}$ and
$R^{(f)}$ is the scalar curvature for $f_{\mu \nu}$.
$M_\mathrm{eff}$ is defined by
\be
\label{Meff}
\frac{1}{M_\mathrm{eff}^2} = \frac{1}{M_g^2} + \frac{1}{M_f^2}\, .
\ee
Furthermore, tensor $\sqrt{g^{-1} f}$ is defined by the square root of
$g^{\mu\rho} f_{\rho\nu}$, that is,
$\left(\sqrt{g^{-1} f}\right)^\mu_{\ \rho} \left(\sqrt{g^{-1}
f}\right)^\rho_{\ \nu} = g^{\mu\rho} f_{\rho\nu}$.
For general tensor $X^\mu_{\ \nu}$, $e_n(X)$'s are defined by
\begin{align}
\label{ek}
& e_0(X)= 1 \, , \quad
e_1(X)= [X] \, , \quad
e_2(X)= \tfrac{1}{2}([X]^2-[X^2])\, ,\nn
& e_3(X)= \tfrac{1}{6}([X]^3-3[X][X^2]+2[X^3])
\, ,\nn
& e_4(X) =\tfrac{1}{24}([X]^4-6[X]^2[X^2]+3[X^2]^2
+8[X][X^3]-6[X^4])\, ,\nn
& e_k(X) = 0 ~~\mbox{for}~ k>4 \, .
\end{align}
Here $[X]$ expresses the trace of arbitrary tensor
$X^\mu_{\ \nu}$: $[X]=X^\mu_{\ \mu}$.

In order to construct the consistent $F(R)$ bigravity,
we add the following terms to the action (\ref{bimetric}):
\begin{align}
\label{Fbi1}
S_\varphi =& - M_g^2 \int d^4 x \sqrt{-\det g}
\left\{ \frac{3}{2} g^{\mu\nu} \partial_\mu \varphi \partial_\nu \varphi
+ V(\varphi) \right\} \nn
& + \int d^4 x \mathcal{L}_\mathrm{matter}
\left( \e^{\varphi} g_{\mu\nu}, \Phi_i \right)\, ,\\
\label{Fbi7b}
S_\xi =& - M_f^2 \int d^4 x \sqrt{-\det f}
\left\{ \frac{3}{2} f^{\mu\nu} \partial_\mu \xi \partial_\nu \xi
+ U(\xi) \right\} \, .
\end{align}
By the conformal transformations
$g_{\mu\nu} \to \e^{-\varphi} g^{\mathrm{J}}_{\mu\nu}$ and
$f_{\mu\nu}\to \e^{-\xi} f^{\mathrm{J}}_{\mu\nu}$,
the total action $S_{F} = S_\mathrm{bi} + S_\varphi + S_\xi$
is transformed as
\begin{align}
\label{FF1}
S_{F} =& M_f^2\int d^4x\sqrt{-\det f^{\mathrm{J}}}\,
\left\{ \e^{-\xi} R^{\mathrm{J}(f)} - \e^{-2\xi} U(\xi) \right\} \nn
& +2m^2 M_\mathrm{eff}^2 \int d^4x\sqrt{-\det g^{\mathrm{J}}}\sum_{n=0}^{4}
\beta_n
\e^{\left(\frac{n}{2} -2 \right)\varphi - \frac{n}{2}\xi} e_n
\left(\sqrt{{g^{\mathrm{J}}}^{-1} f^{\mathrm{J}}} \right) \nn
& + M_g^2 \int d^4 x \sqrt{-\det g^{\mathrm{J}}}
\left\{ \e^{-\varphi} R^{\mathrm{J}(g)} - \e^{-2\varphi} V(\varphi) \right\}
\nn
&+ \int d^4 x \mathcal{L}_\mathrm{matter}
\left( g^{\mathrm{J}}_{\mu\nu}, \Phi_i \right)\, .
\end{align}
The kinetic terms for $\varphi$ and $\xi$ vanish. By the variations
with respect to $\varphi$ and $\xi$ as in the case of convenient $F(R)$
gravity \cite{Nojiri:2003ft}, we obtain
\begin{align}
\label{FF2}
0 =& 2m^2 M_\mathrm{eff}^2 \sum_{n=0}^{4} \beta_n \left(\frac{n}{2} -2 \right)
\e^{\left(\frac{n}{2} -2 \right)\varphi - \frac{n}{2}\xi} e_n
\left(\sqrt{{g^{\mathrm{J}}}^{-1} f^{\mathrm{J}}}\right) \nn
& + M_g^2 \left\{ - \e^{-\varphi} R^{\mathrm{J}(g)} + 2 \e^{-2\varphi}
V(\varphi)
+ \e^{-2\varphi} V'(\varphi) \right\}\, ,\\
\label{FF3}
0 =& - 2m^2 M_\mathrm{eff}^2 \sum_{n=0}^{4} \frac{\beta_n n}{2}
\e^{\left(\frac{n}{2} -2 \right)\varphi - \frac{n}{2}\xi} e_n
\left(\sqrt{{g^{\mathrm{J}}}^{-1} f^{\mathrm{J}}}\right) \nn
& + M_f^2 \left\{ - \e^{-\xi} R^{\mathrm{J}(f)} + 2 \e^{-2\xi} U(\xi)
+ \e^{-2\xi} U'(\xi) \right\}\, .
\end{align}
The Eqs.~(\ref{FF2}) and (\ref{FF3}) can be solved algebraically
with respect to $\varphi$ and $\xi$ as
$\varphi = \varphi \left( R^{\mathrm{J}(g)}, R^{\mathrm{J}(f)},
e_n \left(\sqrt{{g^{\mathrm{J}}}^{-1} f^{\mathrm{J}}}\right)
\right)$ and
$\xi = \xi \left( R^{\mathrm{J}(g)}, R^{\mathrm{J}(f)},
e_n \left(\sqrt{{g^{\mathrm{J}}}^{-1}
f^{\mathrm{J}}}\right) \right)$.
Substituting above $\varphi$ and $\xi$ into (\ref{FF1}),
one gets $F(R)$ bigravity:
\begin{align}
\label{FF4}
& S_{F} = M_f^2\int d^4x\sqrt{-\det f^{\mathrm{J}}}
F^{(f)}\left( R^{\mathrm{J}(g)}, R^{\mathrm{J}(f)},
e_n \left(\sqrt{{g^{\mathrm{J}}}^{-1} f^{\mathrm{J}}}\right) \right) \nn
& +2m^2 M_\mathrm{eff}^2 \int d^4x\sqrt{-\det g}\sum_{n=0}^{4} \beta_n
\e^{\left(\frac{n}{2} -2 \right)
\varphi\left( R^{\mathrm{J}(g)},
e_n \left(\sqrt{{g^{\mathrm{J}}}^{-1} f^{\mathrm{J}}}\right) \right)}
e_n \left(\sqrt{{g^{\mathrm{J}}}^{-1} f^{\mathrm{J}}} \right) \nn
& + M_g^2 \int d^4 x \sqrt{-\det g^{\mathrm{J}}}
F^{\mathrm{J}(g)}\left( R^{\mathrm{J}(g)}, R^{\mathrm{J}(f)},
e_n \left(\sqrt{{g^{\mathrm{J}}}^{-1} f^{\mathrm{J}}}\right) \right) \nn
& + \int d^4 x \mathcal{L}_\mathrm{matter}
\left( g^{\mathrm{J}}_{\mu\nu}, \Phi_i \right)\, , \\
\label{FF4BBB}
& F^{\mathrm{J}(g)}\left( R^{\mathrm{J}(g)}, R^{\mathrm{J}(f)},
e_n \left(\sqrt{{g^{\mathrm{J}}}^{-1} f^{\mathrm{J}}}\right) \right)
\equiv
\left\{ \e^{-\varphi\left( R^{\mathrm{J}(g)}, R^{\mathrm{J}(f)},
e_n \left(\sqrt{{g^{\mathrm{J}}}^{-1} f^{\mathrm{J}}}\right)
\right)} R^{\mathrm{J}(g)} \right. \nn & \left.
  - \e^{-2\varphi\left( R^{\mathrm{J}(g)}, R^{\mathrm{J}(f)},
e_n \left(\sqrt{{g^{\mathrm{J}}}^{-1} f^{\mathrm{J}}}\right)
\right)}
V \left(\varphi\left( R^{\mathrm{J}(g)}, R^{\mathrm{J}(f)},
e_n \left(\sqrt{{g^{\mathrm{J}}}^{-1} f^{\mathrm{J}}}\right)
\right)\right) \right\} \, ,\nn
& F^{(f)}\left( R^{\mathrm{J}(g)}, R^{\mathrm{J}(f)},
e_n \left(\sqrt{{g^{\mathrm{J}}}^{-1} f^{\mathrm{J}}}\right) \right)
\equiv
\left\{ \e^{-\xi\left( R^{\mathrm{J}(g)}, R^{\mathrm{J}(f)},
e_n \left(\sqrt{{g^{\mathrm{J}}}^{-1} f^{\mathrm{J}}}\right)
\right)} R^{\mathrm{J}(f)} \right. \nn
& \left.
  - \e^{-2\xi\left( R^{\mathrm{J}(g)}, R^{\mathrm{J}(f)},
e_n \left(\sqrt{{g^{\mathrm{J}}}^{-1} f^{\mathrm{J}}}\right) \right)}
U \left(\xi\left( R^{\mathrm{J}(g)}, R^{\mathrm{J}(f)},
e_n \left(\sqrt{{g^{\mathrm{J}}}^{-1} f^{\mathrm{J}}}\right)
\right)\right) \right\} \, .
\end{align}
Note that it is difficult to solve Eqs.~(\ref{FF2}) and
(\ref{FF3}) with respect to $\varphi$ and $\xi$ explicitly.
Therefore, it might be easier to define the model
in terms of the auxiliary scalars $\varphi$ and $\xi$ as in (\ref{FF1}).

\subsection{Cosmological Reconstruction and Cosmic Acceleration\label{SIII}}

Let us consider the cosmological reconstruction program
following Ref.~\cite{Nojiri:2012zu} but in slightly extended form
as in \cite{Nojiri:2012re}.

For simplicity, we start from the minimal case
\begin{align}
\label{bimetric2}
S_\mathrm{bi} =&M_g^2\int d^4x\sqrt{-\det g}\,R^{(g)}+M_f^2\int d^4x
\sqrt{-\det f}\,R^{(f)} \nonumber \\
&+2m^2 M_\mathrm{eff}^2 \int d^4x\sqrt{-\det g} \left( 3 - \tr \sqrt{g^{-1} f}
+ \det \sqrt{g^{-1} f} \right)\, .
\end{align}
In order to evaluate $\delta \sqrt{g^{-1} f}$, two matrices $M$ and
$N$, which satisfy the relation $M^2=N$ are taken.
Since $\delta M M + M \delta M = \delta N$, one finds
\be
\label{Fbi7}
\tr \delta M = \frac{1}{2} \tr \left( M^{-1} \delta N \right)\, .
\ee
For a while, we consider the Einstein frame action (\ref{bimetric2}) with
(\ref{Fbi1}) and (\ref{Fbi7b}) but matter contribution is neglected.
Then by the variation over $g_{\mu\nu}$, we obtain
\begin{align}
\label{Fbi8}
0 =& M_g^2 \left( \frac{1}{2} g_{\mu\nu} R^{(g)} - R^{(g)}_{\mu\nu} \right)
+ m^2 M_\mathrm{eff}^2 \left\{ g_{\mu\nu} \left( 3 - \tr \sqrt{g^{-1} f}
\right) \right. \nn
& \left. + \frac{1}{2} f_{\mu\rho} \left( \sqrt{ g^{-1} f } \right)^{-1\,
\rho}_{\qquad
\nu}
+ \frac{1}{2} f_{\nu\rho} \left( \sqrt{ g^{-1} f } \right)^{-1\, \rho}_{\qquad
\mu}
\right\} \nn
& + M_g^2 \left[ \frac{1}{2} \left( \frac{3}{2} g^{\rho\sigma} \partial_\rho
\varphi \partial_\sigma \varphi
+ V (\varphi) \right) g_{\mu\nu} - \frac{3}{2}
\partial_\mu \varphi \partial_\nu \varphi \right] \, .
\end{align}
On the other hand, by the variation over $f_{\mu\nu}$, we get
\begin{align}
\label{Fbi9}
0 =& M_f^2 \left( \frac{1}{2} f_{\mu\nu} R^{(f)} - R^{(f)}_{\mu\nu} \right) \
+ m^2 M_\mathrm{eff}^2 \sqrt{ \det \left(f^{-1}g\right) } \left \{
  - \frac{1}{2}f_{\mu\rho} \left( \sqrt{g^{-1} f} \right)^{\rho}_{\ \nu} \right.
\nn
& \left. - \frac{1}{2}f_{\nu\rho} \left( \sqrt{g^{-1} f} \right)^{\rho}_{\
\mu}
+ \det \left( \sqrt{g^{-1} f} \right) f_{\mu\nu} \right\} \nn
& + M_f^2 \left[ \frac{1}{2} \left( \frac{3}{2} f^{\rho\sigma} \partial_\rho
\xi \partial_\sigma \xi
+ U (\xi) \right) f_{\mu\nu} - \frac{3}{2} \partial_\mu \xi \partial_\nu \xi
\right] \, .
\end{align}
We should note that $\det \sqrt{g} \det \sqrt{g^{-1} f } \neq \sqrt{ \det f}$
in general.
The variations of the scalar fields $\varphi$ and $\xi$ are given by
\be
\label{scalareq}
0 = - 3 \Box_g \varphi + V' (\varphi) \, ,\quad
0 = - 3 \Box_f \xi + U' (\xi) \, .
\ee
Here $\Box_g$ ($\Box_f$) is the d'Alembertian with respect to the metric $g$
($f$).
By multiplying the covariant derivative $\nabla_g^\mu$ with respect to the
metric $g$ with
Eq.~(\ref{Fbi8}) and using the Bianchi identity
$0=\nabla_g^\mu\left( \frac{1}{2} g_{\mu\nu} R^{(g)} - R^{(g)}_{\mu\nu}
\right)$ and
Eq.~(\ref{scalareq}), we obtain
\begin{align}
\label{identity1}
0 =& - g_{\mu\nu} \nabla_g^\mu \left( \tr \sqrt{g^{-1} f} \right)
+ \frac{1}{2} \nabla_g^\mu \left\{ f_{\mu\rho} \left( \sqrt{ g^{-1} f }
\right)^{-1\, \rho}_{\qquad \nu} \right. \nn
& \left. + f_{\nu\rho} \left( \sqrt{ g^{-1} f } \right)^{-1\, \rho}_{\qquad
\mu}
\right\} \, .
\end{align}
Similarly by using the covariant derivative $\nabla_f^\mu$ with respect to the
metric $f$, from (\ref{Fbi9}),
we obtain
\begin{align}
\label{identity2}
0 =& \nabla_f^\mu \left[
\sqrt{ \det \left(f^{-1}g\right) } \left \{
  - \frac{1}{2}\left( \sqrt{g^{-1} f} \right)^{ -1 \nu}_{\ \ \ \ \ \sigma}
g^{\sigma\mu}
  - \frac{1}{2}\left( \sqrt{g^{-1} f} \right)^{ -1 \mu}_{\ \ \ \ \sigma}
g^{\sigma\nu} \right. \right. \nn
& \left. \left.
+ \det \left( \sqrt{g^{-1} f} \right) f^{\mu\nu} \right\} \right]\, .
\end{align}
In case of the Einstein gravity, the conservation law of the energy-momentum
tensor depends from the Einstein equation. It can be derived from the Bianchi
identity.
In case of bigravity, however, the conservation laws of the energy-momentum
tensor of the scalar fields are derived
from the scalar field equations. These conservation laws are independent of the
Einstein equation. The
Bianchi identities give equations (\ref{identity1}) and (\ref{identity2})
independent of the Einstein equation.

We now assume the FRW universes for the metrics $g_{\mu\nu}$ and $f_{\mu\nu}$
and use the conformal time $t$ for the universe with metric $g_{\mu\nu}$
\footnote{
In Ref.~\cite{Nojiri:2012zu}, we have used the cosmological time instead of the
conformal time. The use of the conformal time simplifies the formulation.}:
\begin{align}
\label{Fbi10}
ds_g^2 =& \sum_{\mu,\nu=0}^3 g_{\mu\nu} dx^\mu dx^\nu
= a(t)^2 \left( - dt^2 + \sum_{i=1}^3 \left( dx^i \right)^2\right) \, ,\nn
ds_f^2 =& \sum_{\mu,\nu=0}^3 f_{\mu\nu} dx^\mu dx^\nu
= - c(t)^2 dt^2 + b(t)^2 \sum_{i=1}^3 \left( dx^i \right)^2 \, .
\end{align}
Then $(t,t)$ component of (\ref{Fbi8}) gives
\be
\label{Fbi11}
0 = - 3 M_g^2 H^2 - 3 m^2 M_\mathrm{eff}^2
\left( a^2 - ab \right) + \left(
\frac{3}{4}
{\dot\varphi}^2
+ \frac{1}{2} V (\varphi) a(t)^2 \right) M_g^2 \, ,
\ee
and $(i,j)$ components give
\begin{align}
\label{Fbi12}
0 =& M_g^2 \left( 2 \dot H + H^2 \right)
+ m^2 M_\mathrm{eff}^2 \left( 3a^2 - 2ab - ac \right) \nn
& + \left(
\frac{3}{4} {\dot\varphi}^2 - \frac{1}{2} V (\varphi)
a(t)^2 \right) M_g^2 \, .
\end{align}
Here $H=\dot a / a$.
On the other hand, $(t,t)$ component of (\ref{Fbi9}) gives
\be
\label{Fbi13}
0 = - 3 M_f^2 K^2 + m^2 M_\mathrm{eff}^2 c^2
\left ( 1 - \frac{a^3}{b^3} \right ) + \left(
\frac{3}{4} {\dot\xi}^2 - \frac{1}{2} U (\xi) c(t)^2 \right) M_f^2 \, ,
\ee
and $(i,j)$ components give
\begin{align}
\label{Fbi14}
0 =& M_f^2
\left( 2 \dot K + 3 K^2 - 2 LK \right)
+ m^2 M_\mathrm{eff}^2 \left( \frac{a^3c}{b^2} - c^2
\right) \nn
& + \left(
\frac{3}{4} {\dot\xi}^2 - \frac{1}{2} U (\xi) c(t)^2 \right) M_f^2 \, .
\end{align}
Here $K =\dot b / b$ and $L= \dot c / c$.
Both of Eq.~(\ref{identity1}) and Eq.~(\ref{identity2}) give the identical
equation:
\be
\label{identity3}
cH = bK\ \mbox{or}\
\frac{c\dot a}{a} = \dot b\, .
\ee
If $\dot a \neq 0$, we obtain $c= a\dot b / \dot a$.
On the other hand, if $\dot a = 0$, we find $\dot b=0$, that is, $a$ and $b$
are constant and $c$ can be arbitrary.

We now redefine scalars as $\varphi=\varphi(\eta)$ and
$\xi = \xi (\zeta)$ and
identify $\eta$ and $\zeta$ with the conformal time $t$, $\eta=\zeta=t$.
Hence, one gets
\begin{align}
\label{Fbi19}
\omega(t) M_g^2 =& -4M_g^2
\left ( \dot{H}-H^2 \right )-2m^2 M^2_\mathrm{eff}(ab-ac)
\, , \\
\label{Fbi20}
\tilde V (t) a(t)^2 M_g^2 =&
M_g^2 \left (2 \dot{H}+4 H^2 \right ) +m^2 M^2_\mathrm{eff}(6a^2-5ab-ac)
\, , \\
\label{Fbi21}
\sigma(t) M_f^2 =& - 4 M_f^2 \left ( \dot{K} - LK \right ) - 2m^2 
M_\mathrm{eff}^2 \left ( - \frac{c}{b} + 1 \right ) \frac{a^3c}{b^2}
\, , \\
\label{Fbi22}
\tilde U (t) c(t)^2 M_f^2 =&
M_f^2 \left ( 2 \dot{K} + 6 K^2 -2 L K \right )
+ m^2 M_\mathrm{eff}^2 \left( \frac{a^3c}{b^2} - 2 c^2 + \frac{a^3c^2}{b^3}
\right)
\, .
\end{align}
Here
\be
\label{Fbi23}
\omega(\eta) = 3 \varphi'(\eta)^2 \, ,\quad
\tilde V(\eta) = V\left( \varphi\left(\eta\right) \right)\, ,\quad
\sigma(\zeta) = 3 \xi'(\zeta)^2 \, ,\quad
\tilde U(\zeta) = U \left( \xi \left(\zeta\right) \right) \, .
\ee
Therefore for arbitrary $a(t)$, $b(t)$, and $c(t)$ if we choose $\omega(t)$,
$\tilde V(t)$, $\sigma(t)$, and $\tilde U(t)$
to satisfy Eqs.~(\ref{Fbi19}-\ref{Fbi22}), the cosmological model with given
$a(t)$, $b(t)$ and $c(t)$ evolution can be reconstructed.
Following this technique we presented number of inflationary and/or dark energy
models as well as unified inflation-dark energy cosmologies in above papers.
The method is general and may be applied to more exotic and more complicated
cosmological solutions.

\section{Discussion}

In summary, we revisited the issue of accelerating early-time and/or 
late-time universe in frames of modified gravity. Specifically, the 
following theories were discussed: convenient $F(R)$ and $F(G)$ gravities 
and string-inspired scalar-Einstein-Gauss-Bonnet theory. Scalar-tensor and 
fluid representations of such theories are derived. Working with the FRW-like 
equations we demonstrated how the simplest accelerating cosmology emerges 
from modified gravity. The reconstruction program which gives the 
possibility to derive the requested universe evolution within specific 
modified gravity is developed. The realization of dark energy universe is 
discussed in detail for several models. It is remarkable that large number 
of modified gravity models are viable and may pass the observational 
bounds (for recent discussion, see \cite{Clemson:2012im,Bamba:2012qi}).

As some extension, we formulated the massive $F(R)$ bigravity which is 
free of massive ghost. Its scalar-tensor presentation turns out to be the 
most convenient description of the theory. The presence of not only 
physical metric but also reference metric is the qualitative feature of 
such bigravity. The reconstruction program within massive $F(R)$ bigravity 
is also developed. It gives the possibility to realize the accelerating 
cosmology in terms of massive bigravity.

Number of important questions should be still addressed.
First of all, more precise observational bounds may indicate towards to 
the most realistic modified gravities. In this respect, the perturbations 
theory which is not yet  satisfactory understood in modified gravity 
requests 
a lot of attention. From the other side, it is possible that most 
interesting modified gravity is not yet explored. Thus, the hunt for 
viable modified gravity should continue.

\section*{Acknowledgments}

SDO is grateful to A. Borowiec and M. Francaviglia for kind invitation to 
deliver this short course of lectures at Karpacz Winter School 2013.
The work by SN is supported in part by the JSPS Grant-in-Aid for Scientific
Research (S) \# 22224003 and (C) \# 23540296.
The work by SDO is supported in part by MINECO (Spain), project
FIS2010-15640, by AGAUR (Generalitat de Ca\-ta\-lu\-nya),
contract 2009SGR-994 and by MES project 2.1839.2011 (Russia).

\end{document}